\documentclass[superscriptaddress,aps,pra,twocolumn,showpacs,nofootinbib,longbibliography]{revtex4-1}
\usepackage{amsmath,amssymb,amsthm}
\usepackage{easybmat}
\usepackage[colorlinks=true,citecolor=blue,urlcolor=blue]{hyperref}
\usepackage[pdftex]{graphicx}
\usepackage{times,txfonts}
\usepackage{braket}
\usepackage{color}
\usepackage{natbib}

\newcommand{\be}{\begin{equation}}
\newcommand{\ee}{\end{equation}}
\newcommand{\ba}{\begin{eqnarray}}
\newcommand{\ea}{\end{eqnarray}}

\newcommand{\tr}{\operatorname{Tr}}

\begin{document}
\title{Remote State Preparation using Correlations beyond Discord}  

\author{Som Kanjilal} \email{somkanji@jcbose.ac.in}
  \author{Aiman Khan}

\affiliation{Center for Astroparticle
  Physics and Space Science (CAPSS),  Bose Institute, Block EN, Sector
  V,  Salt Lake,  Kolkata 700  091, India} 
 \author{C.  Jebarathinam} \email{jebarathinam@bose.res.in}
\affiliation{S. N. Bose National Centre for Basic Sciences, Salt Lake,
  Kolkata      700     106,      India}     \author{Dipankar Home}\email{quantumhome80@gmail.com} \affiliation{Center for Astroparticle
  Physics and Space Science (CAPSS),  Bose Institute, Block EN, Sector
  V,  Salt Lake,  Kolkata 700  091, India}

\date{\today}
\begin{abstract}
In recent years, exploring the possible use of separable states as resource for achieving quantum information processing(QIP) tasks has been gaining increasing significance. In this context, a particularly important demonstration has been that non-vanishing discord is the necessary condition for the separable states to be used as resource for remotely preparing any arbitrary pure target state [Nature Physics $8$, $666(2012)$]. The present work stems from our observation that not only resource states with same discord can imply different efficiencies (in terms of average fidelity) of the remote state preparation (RSP) protocol, but also states with higher discord can imply lower RSP efficiency. This, therefore, necessitates identification of the relevant feature of quantum correlations which can appropriately quantify effectiveness of the resource state for the RSP protocol. To this end, for the two-qubit Bell-diagonal states, we show that an appropriate measure of simultaneous correlations in three mutually unbiased bases can serve to quantify usefulness of the resource for the RSP task using entangled as well as separable states, including non-discordant states as resource. In particular, it is revealed that zero-discord states having such non-vanishing measure can be useful for remotely preparing a subset of pure target states. Thus, this work shows that, using separable states, an effective resource for QIP tasks such as RSP can be provided by simultaneous correlations in mutually unbiased bases. 
\end{abstract}


\maketitle

\section{Introduction}
The study of far-reaching implications of quantum foundational aspects of entanglement and its various applications in quantum information constitutes one of the most vibrant areas of research in contemporary science. Neverthless, it has been recognized that the paradigm of entanglement captures only a particular segment of correlations inherent in the quantum regime, and the study of useful correlations inherent in separable states, along with their applicability in quantum information processing(QIP) tasks, is attracting increasing attention. This is essentially because, compared to entangled states, separable states are easier to produce, manipulate and protect against decohering effects. In this context, various measures of ''quantumness" of correlations beyond entanglement have been suggested \cite{WH01,GW14,BC+14}, chief amongst them being quantum discord\cite{HV01,WH01}. It is also important to note that while entanglement has been established as a resource for QIP protocols using pure states \cite{BBP+93,BW92}, there exist examples of QIP tasks where bipartite separable states can act as resource. For instance, consider the protocol of quantum state merging where, given an unknown quantum state distributed over two systems, the task is to determine how much quantum communication is needed to transfer the complete state to one system \cite{HO+05}. It has been shown \cite{HOW07,MD11} that quantum discord quantifies the usefulness of separable states in the context of this protocol. Further, consider the task of  deterministic quantum computational protocol of efficient estimation of the trace of a unitary matrix, namely, what is known as the "power of one qubit" model; it has been shown that quantum discord can be used as a figure of merit to characterize whether such a protocol is successfully implemented \cite{DSC08}. Girolami et.al. have also shown that quantum discord determines the necessary condition for bipartite qubit state to be useful for showing the so-called interferometric power of quantum states \cite{GS+14}.

Against the above backdrop, focusing on a particular QIP task, viz, remote state preparation (RSP) \cite{Pati00,L00}, the motivation underlying the present paper stems from the fact that while non-vanishing discord has been shown to provide the necessary condition for successful implementation of the RSP task \cite{DLM+12}, we've found that higher discord states do not necessarily imply higher efficiency (in terms of optimal quadratic fidelity) of the RSP protocol with respect to a specific target state. Thus, there is a need for characterizing the efficiency of RSP for a general class of resource states, including entangled as well as separable states, in terms of a suitable measure of quantum correlations. On the other hand, for the analogous QIP task of teleportation, entanglement is deemed to be the necessary condition \cite{BBP+93,CS+17}, as well as it characterizes the efficiency of the protocol \cite{HH+99}. The central result of this paper is that a suitable measure of efficiency of the RSP protocol (say, in terms of average optimal quadratic fidelity) is found to be in direct correspondence with an appropriate measure of simultaneous correlations in three mutually unbiased bases \cite{GW14}, which is non-zero for any non-product state. This, in particular, enables to capture effectiveness of the zero discord states for remotely preparing a class of pure target states. Further, given any two non-zero discord states as resource for remotely preparing a target state, we show that it is the simultaneous correlation in mutually unbiased bases that determines which particular resource state can be more useful than the other. Explanation of the reason for discord to fail as the quantifier of RSP efficiency is a highlight of the present paper.

The recently introduced measures of simultaneous correlations in mutually unbiased bases(SCMUB) inherent in the quantum state seek to quantify "quantumness" by the persistence of correlations in incompatible bases used to measure the state\cite{GW14,WMC+14}. Correlations in a given basis are quantified using the Holevo quantity, and comparing amongst incompatible bases their respective Holevo quantities, one can obtain a series of quantumness measures by choosing sets of bases that maximize the amount of simultaneous correlations in the sense whose precise meaning will be explained later. Incompatible bases used to define these measures of correlations are chosen to be mutually unbiased, meaning that if a particular member of a basis is measured, it is projected with equal probability onto all members of the corresponding incompatible basis. A class of these measures\cite{GW14} has been shown to vanish if and only if the state in question is a  product state, which in turn raises the question as to whether the correlations captured using such measures can provide an effective characterization of efficiency in QIP tasks using separable states. An affirmative answer to this question is provided in the present paper in the context of a specific QIP task, namely, Remote State Preparation.\\

The plan of the paper is as follows In Section II we give an overview of the general RSP protocol along with the nuances regarding the condition pertaining to successful implementation of the RSP protocol and the role of quantum discord in this regard. This is followed by examples of resource states for which quantum discord cannot successfully explain the higher efficiency of the RSP protocol. In Section III we discuss the effectiveness/strength of the resource state pertaining to RSP protocol in terms of average optimal quadratic fidelity. In Section IV, a suitable measure of simultaneous correlations in mutually unbiased bases is identified and related to the measure of RSP efficiency. In Section V, we take zero discord states as resource and show in what way it can be used as an effective resource for remotely preparing a particular set of pure target states. The ramifications of our results are discussed in Section VI.

\section{Remote State Preparation (RSP)--an overview}
The RSP protocol \cite{Pati00} seeks to demonstrate that by using a shared bipartite state as resource, a specified quantum state (hereby referred to as target state) can be prepared at a distant location by classical communication of local quantum measurement outcomes. Operationally, we can outline the RSP protocol as follows: Alice and Bob share a general bipartite qubit state given by
\begin{equation}
\label{e0}
\rho = \frac{1}{4}\left(\mathcal{I}\otimes\mathcal{I} +\sum_{i}a_{i}\sigma_{i}\otimes\mathcal{I}+\sum_{j}b_{j}\mathcal{I}\otimes \sigma_{j}+\sum_{k}E_{k}\sigma_k\otimes\sigma_k\right)
\end{equation}
here $i,j,k=1,2,3$ and $|E|_{1}\geq |E|_{2}\geq |E|_{3}$ are the non-zero diagonal elements of the correlation matrix E. In this paper special emphasize is given on the class of states for which the local marginals of Alice and Bob are maximally mixed, i.e. $a_{i}=b_{i}=0$ for $i=1,2,3$. This particular class of bipartite qubit state is known as Bell diagonal state. Both parties agree to prepare a pure state(specified, say, by Bloch vector $\vec{t}$) in the plane perpendicular to Bloch vector $\vec{\beta}$. Thus, the set of all target states for a particular RSP protocol correspond to the great circle specified by the unit norm Bloch vector $\vec{\beta}$ perpendicular to the great circle. This implies that different instances of RSP protocols correspond to different Bloch vector $\vec{\beta}$. 

Consider a typical RSP scenario where Alice performs a projective, two-outcome measurement specified by $\vec{\alpha}$. The measurement outcomes $\alpha=\pm1$ are encoded in a single classical-bit and communicated to Bob. Bob then performs a local operation corresponding to Alice's measurement outcome. If $\alpha=1$, Bob performs no operation, while if $\alpha=-1$, a $\pi$ rotation perpendicular to $\vec{\beta}$ is performed on Bob's particle. The Bloch vector of the state prepared as a result of these steps is denoted by $\vec{t}$.\\

It has been shown that if maximally entangled state is used as shared bipartite state, the known target state can be exactly prepared (maximum efficiency) \cite{L00,Pati00}. Furthermore, if one uses noisy entangled state as resource, then also one can implement RSP \cite{XL+05}, albeit with less efficiency. In that case, the actually prepared state is, in general, a mixture of a pure state and its orthogonal state. If the weightage factors of such a mixture are \textit{unequal}, this is deemed to be the criterion for an effective implementation of RSP \cite{XL+05}. This is because, corresponding to a target state, if the optimally prepared state in RSP turns out to be maximally mixed, then in such a scenario, the correlations in the resource state do not play any role. Thus, the signature of what constitutes an effective implementation of RSP protocol is that the prepared state is different from maximally mixed state \cite{DLM+12}. 

Now, considering the  efficiency of an effective  RSP protocol, this is directly proportional to how much closer to the target state (specified by Bloch vector $\vec{t}$) is the prepared state (specified by Bloch vector $\vec{p}$); this can be quantified in terms of either the quadratic fidelity (equivalently, the pay-off function) $\mathcal{P}_{q}(\vec{t},\vec{p}) = (\vec{t}.\vec{p})^2$ or the linear fidelity  $\mathcal{P}_{l}(\vec{t},\vec{p})=\frac{1}{2}(1+\vec{t}.\vec{p})$. If $\mathcal{P}_{q}$ is zero (or equivalently, $\mathcal{P}_{l}$ is half) then the prepared state is maximally mixed; i.e., the RSP protocol has not been effectively implemented. If the prepared state is the same as target state then $\mathcal{P}_{q}=1$ (or, equivalently $\mathcal{P}_{l}=1$) whence RSP is implemented with maximum efficiency. Therefore, we can take both $\mathcal{P}_{q}$ and $\mathcal{P}_{l}$ as measure of efficiency.The treatment that follows in this paper is given in terms of quadratic fidelity. For a particular target state, the maximum efficiency (quadratic fidelity\cite{HT+14}) that can be achieved is given by
\begin{equation}
\label{e1}
\max_{\vec{p}}  \mathcal{P}_{q}\left(\vec{t},\vec{p}\right)=\mathcal{P}_{q}^{max}\left(\vec{t}\right)
\end{equation}
 where the maximization is done over all remotely prepared states that can be achieved with respect to local operations of Alice and $\rho$ is the shared state.
 
 Since a particular RSP protocol is specified by the shared information about the angle $\beta$ of the great circle on the Bloch sphere, the effectiveness of the shared state corresponding to a RSP protocol where the target states lie on the great circle specified by $\beta$ can be given by \cite{DLM+12}
 \begin{equation}
     \label{e2}
     \mathcal{P}_{q}^{av}\left(\vec{\beta}\right)= \frac{\int d\vec{t}\mathcal{P}_{q}^{max}\left(\vec{t},\rho\right)}{2\pi} 
 \end{equation}
 
Here, the integration is performed over all pure target states $\vec{t}$ lies on the great circle $\vec{\beta}$. If $\mathcal{P}_{q}^{av}(\vec{\beta})$ is zero then an effective RSP protocol cannot be implemented for any pure target state that lies on the great circle specified by $\vec{\beta}$. Therefore, $\mathcal{P}_{q}^{av}(\vec{\beta})\neq 0$ gives the necessary condition for an effective implementation of RSP. Note that, here averaging of the maximum RSP efficiencies corresponding to different target states, $\mathcal{P}_{q}^{av}(\vec{\beta})$ characterizes effectiveness of the resource states for the given RSP protocol. Further, if one considers minimization of $\mathcal{P}_{q}^{av}(\vec{\beta})$ with respect to $\beta$ and find the condition on the shared resource state such that the $F_{q}=\min_{\beta}\mathcal{P}_{q}^{av}(\vec{\beta})\neq 0$, then this would give the necessary condition on the resource state such that it can implement RSP protocol corresponding to any great circle.i.e. 

$$\mathcal{P}_{q}^{av}\left(\vec{\beta}\right)\neq 0 \hspace{5mm} \forall 0\leq \beta \leq \pi$$

Now, it is relevant to note the connection between the quantity $F_{q}$ and quantum discord which, for Bell diagonal states, is given by
 \begin{equation}
     \label{e3}
     \mathcal{D}\left(\rho_{AB}\right)=\sqrt{\frac{E_{2}^{2}+E_{3}^{2}}{2}}
 \end{equation}
It is thus seen that $\mathcal{D}(\rho_{AB})$ is equal to the quantity $F_{q}=\min_{\beta}\mathcal{P}_{q}^{av}(\vec{\beta})$ \cite{DLM+12}. Therefore, $F_{q}=\min_{\beta}\mathcal{P}_{q}^{av}(\vec{\beta})\neq 0$ implies non-vanishing quantum discord. Thus, only if quantum discord is non-zero, using Bell diagonal resource state one can effectively implement the RSP protocol for at least some  pure target state corresponding to any great circle in the Bloch sphere. 

From the above discussion it follows that if a state with vanishing discord is used as resource for RSP, the most one can contend is the existence of at least one great circle for which RSP \textit{cannot} be implemented for \textit{any} pure target state on that great circle. Nevertheless, this \textit{does not} rule out using zero discord state as resource for an effective RSP corresponding to at least one great circle. 

Next, let us examine the question of quantification of effectiveness of the resource state for RSP. Recall that, a shared state is considered to be effective for RSP of a given target state $\vec{t}$ if the remotely prepared state is not maximally mixed ($\vec{p} \neq 0$) and is an unequal mixture of the target state
and its orthogonal state; in other words, then the maximum quadratic fidelity $\mathcal{P}_{q}^{max}(\vec{t}) \neq 0$. Quantification of effectiveness of the resource state for remotely preparing a given target state would then entail comparing two different resource states for the purpose of an effective RSP in the following sense: Given two different resource states, one of them is deemed to be more effective than the other for RSP if the prepared state lies further from the maximally mixed state compared to using the other resource state.  

Next, from the feature that quantum discord is equal to the minimum of the average fidelity $\mathcal{P}_{q}^{av}(\vec{\beta})$, it is prima facie evident that discord cannot quantify the effectiveness of the resource state in the RSP protocol for any given $\vec{\beta}$. To be more precise, if we take two separable bipartite quantum states having non-zero discord and ask which one of them can perform better as resource for RSP of an arbitrary target state, it may happen that the state with less quantum discord will perform better. To see this explicitly, consider the following two classes of states, namely, maximally entangled state mixed with correlated and anti-correlated noise respectively.
\begin{equation}
\label{e4}
    \rho_{c}(p)=p\ket{\psi^{-}}\bra{\psi^{-}}+\frac{(1-p)}{2}(\ket{01}\bra{01}+\ket{10}\bra{10})
\end{equation}
\begin{equation}
\label{e5}
    \rho_{g}(p)=p\ket{\psi^{-}}\bra{\psi^{-}}+\frac{(1-p)}{2}(\ket{00}\bra{00}+\ket{11}\bra{11})
\end{equation}

Next, the argument goes as follows. Consider RSP of a particular pure target state, $\ket{\psi(\theta,\phi)}=\cos\theta \ket{0}+e^{i\phi}\sin\theta \ket{1}$. It can be shown that maximum quadratic fidelity $P_{q}^{max}$ corresponding to the resource states $\rho_{c}$ and $\rho_{g}$ are respectively
\begin{equation}
    \label{e6}
    \mathcal{P}_{q}^{max}(\rho_{c}(p))=\frac{1}{2}[1+p+(1-p)\cos 2\theta]
\end{equation}
and 
\begin{equation}
    \label{e7}
    \mathcal{P}_{q}^{max}(\rho_{g}(p))=\frac{1}{2}[3p+(1-p)\cos 2\theta-1]
\end{equation}
It can be easily checked that $\mathcal{P}_{q}(\rho_{c}(p)) \geq \mathcal{P}_{q}(\rho_{g}(p))$ for all values of $p$.
Then comparing the effectiveness of $\rho_{c}(p)$ and $\rho_{g}(p)$ for RSP protocol corresponding to the great circle characterized by $\vec{\beta}$ containing the pure target state $\ket{\psi(\theta,\phi)}$, it is clear from Eqs. (\ref{e6}) and (\ref{e7}) that $\rho_{c}(p)$ outperforms $\rho_{g}(p)$; therefore, 
 $\mathcal{P}_{q}^{av}(\vec{\beta},\rho_{c}(p)) \geq \mathcal{P}_{q}^{av}(\vec{\beta},\rho_{g}(p))$ implying $\rho_{c}(p)$ to be more effective as the resource state compared to $\rho_{g}(p)$ for the RSP protocol characterized by any $\vec{\beta}$.

 Next, note that for $p\leq \frac{1}{3}$, $\mathcal{D}(\rho_{c}) \leq \mathcal{D}(\rho_{g})$, which can be checked from Eqs(\ref{e4}) and (\ref{e5}) using Eq.(\ref{e3}). In view of the preceding discussion it follows that, if $p\leq \frac{1}{3}$, lower discord states $\rho_{c}(p)$ correspond to higher efficiency in the RSP protocol characterized by any $\vec{\beta}$. Note that this result holds good even if one uses maximum linear fidelity [cite] as a measure of efficiency in the RSP protocol characterized by $\beta$.
 
 The above analysis naturally begs the question as to what aspect of quantum correlation can serve as an appropriate quantifier of efficiency of the resource state used for RSP corresponding to a given  $\vec{\beta}$. This is addressed in the following section using the notion of simultaneous correlations in mutually unbiased bases (SCMUB) whose preliminary ideas have been already introduced in the introduction.
 
\section{Quantifying resource for RSP using simultaneous correlations in mutually unbiased bases (SCMUB)}

Let Alice and Bob share a bipartite $d\times d$ dimensional state $\rho_{AB}$.  Upper bound on the accessible information about Alice's measurement $\{\Pi_{i}^{A}:=\ket{a_{i}}\bra{a_{i}}\}$ available to Bob when $\rho_{AB}$ is used as a resource channel is given by
the Holevo quantity pertaining to the ensemble $\{p_{i},\rho^{B}_{i}\}$ corresponding to Bob's end \cite{HV01} and is defined as follows:
\begin{equation}
\chi\left(\rho_{AB},\{\Pi_{i}^{A}\}\right):=S\left(\sum_{i}p_{i}\rho^{B}_{i}\right)- \sum_{i}p_{i}S\left(\rho^{B}_{i}\right),
\end{equation}
where $S(\rho)=-\tr[\rho\log\rho]$ is the von Neumann entropy and  $\rho^{B}_{i}=\tr_{A}(\Pi_{i}^{A} \otimes \openone_B \rho_{AB})/p_{i}$, with $p_{i}=\tr(\Pi_{i}^{A} \otimes \openone_B \rho_{AB})$ and $\openone_B$ is the identity operator on Bob's side.
The maximum accessible information available to Bob about distinguishable outcomes (classical random variable) pertaining to an arbitrary measurement basis $\{\Pi_{i}^{A}\}$  when $\rho_{AB}$ is used as a resource channel is given by
\begin{align}
    \mathcal{C}_{1}   & = \max_{\{\Pi_{i}^{A}\}}\chi\left(\rho_{AB},\{\Pi_{i}^{A}\}\right)
    \label{e9}
\end{align}
 Therefore, one can interpret $\mathcal{C}_{1}$ given by above Eq. (\ref{e9}) as the maximum amount of classical information contained in a state $\rho_{AB}$ \cite{WMC+14}.

A fundamental feature of quantum mechanics is the existence of mutually unbiased bases. Two sets of complete bases, say $\{\ket{a_{i}^{1}}\}$ and $\{\ket{a_{j}^{2}}\}$ 
in Hilbert space of dimension $d$ are defined to be mutually unbiased if and only if 
$$ |\langle a_{i}^{1}|a_{j}^{2}\rangle|=\frac{1}{\sqrt{d}}.$$ 
  
  It was pointed out in \cite{GW14,WMC+14} that simultaneous existence of accessable information in sets of mutually unbiased bases is a quantum phenomena. Based on this idea, a series of measures that seek to capture quantum correlations corresponding to a state $\rho_{AB}$,have been proposed \cite{GW14,WMC+14}. To be more precise, let us denote by $\Omega$ the set of all pairs of bases that are mutually unbiased with each other, that is to say:

\begin{align}
    \Omega_{A} & := \{ \{ \{\ket{a_i^1}_{A}\}, \{\ket{a_j^2}_{A}\} \}: |_{A}\braket{a_i^1|a_j^2}_{A}| = \frac{1}{\sqrt{d}} \nonumber\\
    & \forall i,j \in (1,2,...,d) \} \nonumber
\end{align}

One can now define the quantity $\mathcal{C}_2$ as the maximum amount of simultaneous correlations that exist in any given pair of mutually unbiased bases (SCMUB), that is \cite{GW14}:

\begin{equation} 
\label{e10}
    \mathcal{C}_2 = \max_{\Pi_1^A,\Pi_2^A\in\Omega} \min [\chi(\rho_{AB},\{\Pi_1^A\}),\chi(\rho_{AB},\{\Pi_2^A)\}],
\end{equation}
where $\{\Pi_i^A\}$ represents the basis of measurement in Alice's local Hilbert space. This definition can be easily generalized by performing the minimization process above using more than two mutually unbiased bases pertaining to Alice's part of the joint Hilbert space. For a bipartite quantum state $\rho_{AB}$ with the local Hilbert space dimension in Alice's side being $d$, one can define the quantity as in Eq. (\ref{e10}) with $m$ mutually unbiased bases, here $3 \le m \le d+1$.

Note that for $d=2$, the measure of SCMUB defined as in Eq. (\ref{e10}) cannot be generalized with more than three bases since for qubit systems there cannot be more than three mutually unbiased bases. Next, similar to the quantity $\mathcal{C}_2$, one can define $\mathcal{C}_3$ as follows  \cite{GW14}:

\begin{equation} 
\label{e11}
    \mathcal{C}_3 = \max_{\Pi_1^A,\Pi_2^A,\Pi_3^A\in\Lambda_{A}} \min [\chi(\rho_{AB},\{\Pi_1^A\}),\chi(\rho_{AB},\{\Pi_2^A\}),\chi(\rho_{AB},\{\Pi_3^A\})],
\end{equation}
where the set of all triads of mutually unbiased bases in Alice's Hilbert space is denoted by $\Lambda_{A}$ as
\begin{align}
    \Lambda_{A} & := \{ \{ \{\ket{a_i^1}_{A}\}, \{\ket{a_j^2}_{A}, \{\ket{a_k^3}_{A}\} \}: |_{A}\braket{a_i^1|a_j^2}_{A}| = |_{A}\braket{a_j^2|a_k^3}_{A}|\nonumber\\
    & =|_{A}\braket{a_k^3|a_i^1}_{A}| = \frac{1}{\sqrt{d}} \hspace{5mm} \forall i,j,k \in (1,2,...d)\}.
\end{align}

To date, the quantity $\mathcal{C}_3$ has been explicitly calculated only for certain states. The key results relevant to our purpose here are as follows:

\begin{itemize}
    \item $\mathcal{C}_3(\rho_{AB})=0$ iff $\rho_{AB}$ is a product state. 
    \item For bipartite qubit Bell diagonal states($\vec{a}=\vec{b}=0, E_{jk} = \delta_{jk}\mbox{E}_j$), the quantity $C_{3}$ is given by 
    \begin{equation}
    \label{c3bd}
        \mathcal{C}_3(\rho_{AB}) = 1 - \mbox{h}\left(\frac{1+\sqrt{\left(\mbox{E}_1^2+\mbox{E}_2^2+\mbox{E}_3^2\right)/3})}{2}\right)
    \end{equation}
 where h is the von Neumann entropy function.
\end{itemize}

Now, recalling the earlier discussed example in Section II showing an inadequacy of quantum discord to correctly quantify the efficiency of RSP protocol pertaining to target states on a specified great circle characterized by $\vec{\beta}$, note that from Eqs. (\ref{e6}) and (\ref{e7}) it can be seen that $\mathcal{C}_{3}(\rho_{c})$ is greater than $\mathcal{C}_{3}(\rho_{g})$ corresponding to $\mathcal{P}_{q}^{av}(\beta,\rho_{c})\geq \mathcal{P}_{q}^{av}(\beta,\rho_{g})$, thereby circumventing the inadequacy of quantum discord in this context.

At this stage, it is important to recognize that the above analysis is based on the measure of efficiency $\mathcal{P}_{q}^{av}(\vec{\beta})$ defined by Eq. (\ref{e2}) which compares the effectiveness using two different resource states for a specific RSP protocol which essentially corresponds to a particular $\vec{\beta}$. Thus, for the completeness of the above analysis, it is required to consider the gamut of RSP protocols for a given resource state by spanning the entire range of $\vec{\beta}$ and define a suitable measure of RSP efficiency for that particular resource state as follows: 

\begin{equation}
\label{e13}
    \mathcal{G} = \int \frac{d\vec{\beta}\mathcal{P}_{q}^{av}(\vec{\beta})}{\int d\vec{\beta}}
\end{equation}
 where the integrations in the numerator and denominator are done over all choices of $\vec{\beta}$. Note that if $\mathcal{G}$ vanishes, this means that there are no states that can be remotely prepared closer to the target state other than the maximally mixed state, thereby indicating ineffectiveness of the RSP protocol for a given resource state corresponding to any value of $\vec{\beta}$; i.e., $\mathcal{P}_{q}^{av}(\vec{\beta})=0$ for all $\vec{\beta}$. Therefore, non-vanishing of the quantity $\mathcal{G}$ can be taken to signify a necessary condition for effectiveness of the RSP protocol for a given resource state.

One can now calculate $\mathcal{G}$ given by Eq. (\ref{e13} for Bell-diagonal states by averaging $\mathcal{P}_{q}^{av}(\vec{\beta})$ over all choices of Bloch vector $\vec{\beta}$. This can be done by noting that the optimized quadratic fidelity for a particular RSP scenario, averaged over all target states can be expressed in the following form \cite{DLM+12}:

\begin{equation}
\label{e14}
    \mathcal{P}_{q}^{av}(\vec{\beta}) = ||E||^2 - \vec{\beta}^{\dagger}(E^{\dagger}E)\vec{\beta}
\end{equation}
where $||E||^2 = \sum_{jk}E_{jk}^2 = \sum_i E_i^2$ is the invariant norm of the correlation matrix E for the quantum state and $A^{\dagger}$ is the transpose of the quantity $A$. The Bloch vector $\vec{\beta}$ is now represented as the column vector. The term $\vec{\beta}^{\dagger}(E^{\dagger}E)\vec{\beta}$, when averaged over all choices of $\vec{\beta}$, can be shown to be equal to $\frac{1}{3}\sum_i E_i^2$ (its derivation is given in Appendix A), which, together with Eq. (\ref{e14}), implies the following, expression for the quantity $\mathcal{G}$

\begin{equation}
\label{e15}
    \mathcal{G} = \frac{2}{3}\left(\mbox{E}_1^2 + \mbox{E}_2^2 + \mbox{E}_3^2\right)
\end{equation}

Let us now pose more precisely the issue of effectiveness of resource states for RSP pertaining to any pure target state. Consider a target state $\vec{t}=\{\vec{t}_{i}\}$ with $i=1,2,3$ are the components of the Bloch vector $\vec{t}$ along three mutually orthogonal directions and two Bell diagonal states $\rho$ and $\rho'$ with the correlation matrix $\text{diag}[E_{1},E_{2},E_{2}]$ and $\text{diag}[E_{1}',E_{2}',E_{2}']$ respectively.  Let both $\rho$ and $\rho'$ be able to remotely prepare the target state as non-maximally mixed state. If we assume $\rho$ outperforms $\rho'$ for such a target state, then the following should hold good
\begin{equation}
    \mathcal{P}_{q}^{max}(\rho,\vec{t}) \geq \mathcal{P}_{q}^{max}(\rho',\vec{t})\nonumber
\end{equation}
whence,
\begin{equation}
    \sum_{i} E_{i}^{2} t_{i}^{2} \geq \sum_{i} {E'}_{i}^{2} t_{i}^{2} \nonumber
\end{equation}
and
\begin{equation}
    \label{e16}
\sum_{i}[E_{i}^{2}-{E'}_{i}^{2}]t_{i}^{2} \geq 0
\end{equation}

Note that 
\begin{equation}
\sum_{i}[E_{i}^{2}-{E'}_{i}^{2}]\geq \sum_{i}[E_{i}^{2}-{E'}_{i}^{2}]t_{i}^{2} \nonumber
\end{equation}
therefore
\begin{equation}
    \label{e17}
   \sum_{i}E_{i}^{2} \geq \sum_{i}{E'}^2_{i}
\end{equation}
From Eq. (\ref{e17}), using Eq. (\ref{e15}), one can infer $\mathcal{G}(\rho)\geq\mathcal{G}(\rho')$. Therefore, given a target state and two resource states, the RSP corresponding to a higher value of $\mathcal{G}$ works better. Thus, the quantity $\mathcal{G}$ can serve as an appropriate measure of effectiveness of the shared state corresponding to RSP protocol as explained above. 

Notice that the fidelity quantity $\mathcal{G}$ does not vanish for zero-discord Bell-diagonal states (which are of the form $E=\text{diag}[E_{1},0,0])$, showing that zero discord states may still be of use for remote preparation of at least some target state since  the optimized quadratic fidelity $\mathcal{P}_{q}^{max}$ has support on the set of target states defined by at least some choice of the Bloch vector $\vec{\beta}$. In other words, zero-discord states are useful as resource for implementing RSP for a class of target states pertaining to at least one choice of Bloch plane perpendicular to $\vec{\beta}$. Thus, this complements the results of Dakic et al \cite{DLM+12}, who showed that for all discordant states, RSP can be implemented for pure target states drawn from \textit{any} great circle corresponding to an arbitrary choice of $\vec{\beta}$. 

\section{Relating the fidelity quantity $\mathcal{G}$ with the SCMUB measures}

 Having defined in the preceding section an appropriate fidelity quantity $\mathcal{G}$ for RSP, in order to connect it with a suitable measure of correlations in the resource state,  we obtain from Eqs. (\ref{c3bd}) and (\ref{e15}) the following analytical relationship between $C_{3}$ and $\mathcal{G}$ pertaining to two-qubit Bell diagonal states:
    
    \begin{equation} 
        \label{C3G}
        \mathcal{C}_3 = 1 - \mbox{h}\left(\frac{1+\sqrt{\mathcal{G}}}{2\sqrt{2}}\right)
        \end{equation}
    It can be shown that $\mathcal{C}_{3}$ is a monotonically increasing function of $\mathcal{G}$. To see this, note that, if a function $f(x)$ is continuous and differentiable in an arbitrary interval $a \leq x \leq b$ then $f(x)$ is a monotonically increasing function of $x$.  Now, if the derivative of $f(x)$ with respect to $x$ is not negative, i.e., $\frac{df(x)}{dx} \geq 0$ for all $x \in [a,b]$, by differentiating the left hand side of Eq. (\ref{C3G}) with respect to $\mathcal{G}$, it  can be checked that $\frac{d\mathcal{C}_{3}}{d\mathcal{G}} \geq 0$ for all values of $\mathcal{G}$ which implies that as $\mathcal{G}$ increases, $\mathcal{C}_3$ also increases. 
    
Upshot of the above argument is that the relationship between $C_3$ and $\mathcal{G}$ given by Eq. (\ref{C3G}) implies that for any pair of two-qubit Bell diagonal states, higher value of $C_3$ always implies higher value of $\mathcal{G}$, which in turn implies higher value of efficiency of the resource state for RSP. Therefore, for this class of states, it is legitimate to regard the simultaneous correlation in three mutually unbiased bases, quantified by $C_{3}$, as quantitative resource for the RSP protocol characterized by the quantity $\mathcal{G}$.
 
It then immediately follows that the vanishing of $\mathcal{G}$ necessarily implies the vanishing of $C_{3}$ which, as mentioned earlier, corresponds to product states. We can therefore infer that for RSP using product states, the fidelity function has no support on any plane defined by the direction $\vec{\beta}$, that is to say, using the RSP protocol, no state can be remotely prepared closer to the target state than the maximally mixed state. This implies that all non-product resource states (including zero-discord states) can be used to remotely prepare some fraction of states that are closer to the target state than the maximally mixed state.  

Here it is relevant to note that Horodecki et. al.\cite{HT+14} discussed the efficiency of the RSP protocol in terms of the minimization of average fidelity (quadratic or linear) over all the great circles. However, such a quantity as the efficiency measure does not capture the feature mentioned above about usefulness of any non-product state as resource in RSP pertaining to at least some fraction of target states. In other words, the existence of any great circle for which RSP can be non-trivially effective can only be inferred from the averaging of $\mathcal{F}_{q}^{av}(\vec{\beta})$ over all the great circles characterized by $\vec{\beta}$, as has been shown in our above discussion.  

Finally, we take note of the work by Giorgi \cite{G13} which showed that using local operations one can induce quantum discord in the shared resource state of initially zero discord, thereby making the resource state effective for RSP protocol corresponding to any great circle of the Bloch sphere, denoted by $\vec{\beta}$. In light of the treatment given in this paper, it seems enticing to try to investigate whether the validity of the aforementioned result can be shown by appropriately viewing the local operations as physically relevant unitary operations.

\section{Summary and Outlook}

To put it succinctly, the measure of simultaneous correlations in three mutually unbiased bases denoted as $C_{3}$ has been demonstrated to be the appropriate quantifier of efficiency for  effectively implementing remote preparation of a non-vanishing fraction of target states which, in turn, leads to the implication that zero discord states can also serve as an effective resource state for RSP. This finding is distinct from the one made in \cite{DLM+12}, where the necessity of quantum discord has been argued to ensure that there is a finite proportion of target states that can be remotely prepared for \textit{any} choice of Bloch vector $\vec{\beta}$. Relaxing the efficiency requirement so as to entail not the worst case scenario but rather the average effect, a significant consequence is that correlations inherent in the resource state beyond discord can become useful in enabling meaningful remote preparation of a finite proportion of target states pertaining to at least some great circle on the Bloch sphere.  Moreover, in the treatment by Dakic et. al. \cite{DLM+12}, the characterization of RSP, which is in the worst case scenario, involves the correlations of the two-qubit states with respect to the Bell-diagonal state along only any two axes. On the other hand, in general, the implementation of RSP makes use of the correlations present in the resource state pertaining to all three axes. Thus, the quantity $C_{3}$ which is a function of correlations in all three axes of the two-qubit state becomes naturally most suited for characterizing the role of the resource state for RSP.

Here we may stress that this work constitutes the first application of simultaneous correlation in mutually unbiased bases(SCMUB) in the context of a QIP task, thereby complementing the recent demonstration of quantitative relationship between the amount of steerability and the SCMUB measures \cite{JKK+18}. Thus, such investigations underscore the need to comprehensively explore various ramifications of SCMUB in the context of both quantum fundamental effects and QIP tasks.

\textit{Acknowledgements:-} CJ thanks A.R.P Rau and U Sinha for helpful discussions in the initial stage of this work during his visit to Raman Research Institute, Bangalore. SK acknowledges the support of a fellowship from the Bose Institute, Kolkata. CJ acknowledges S. N. Bose Centre, Kolkata for the postdoctoral fellowship.

\bibliography{biblio}

\appendix
\section{Supplementary}
Here we will show the following result for the two-qubit Bell diagonal state
\begin{equation}
    \label{s1}
    \frac{\int d\vec{\beta}\vec{\beta}^{\dagger}(E^{\dagger}E)\vec{\beta}}{d\vec{\beta}}=\frac{1}{3}||E||^{2}
\end{equation}
Let us take 
\begin{equation}
    \label{s2}
\vec{\beta}=\begin{bmatrix}
\sin\theta\sin\phi\\
\sin\theta\cos\phi\\
\cos\theta
\end{bmatrix}
\end{equation}
Then, the L.H.S of Eq. (\ref{s1}) can be written as
\begin{equation}
\frac{\int\int [E_{1}^{2}\sin^{2}\theta\sin^{2}\phi+E_{2}^{2}\sin^{2}\theta\cos^{2}\phi+E_{1}^{2}\cos^{2}\theta]\sin\theta d\theta d\phi}{\int\int sin\theta d\theta d\phi} \nonumber
\end{equation}

Note that, Bloch vectors perpendicular to all the great circles lie on one hemisphere of the Bloch sphere. Therefore, if we start from the great circle corresponding to the equator and continue to vary the orientation of the great circle to get the different values of $\theta$ and $\phi$, the limits of the integration will be from $0$ to $\pi$ for $\theta$ and $0$ to $\pi$ for $\phi$. 
Then, using the above mathematical expression for LHS pertaining Eq. (\ref{s1}) and putting the corresponding limits in the integration we obtain from Eq. (\ref{s1}) the following result used in the text preceding Eq. (\ref{e15})
\begin{equation}
   \label{s2}
    \frac{\int d\vec{\beta}\vec{\beta}^{\dagger}(E^{\dagger}E)\vec{\beta}}{d\vec{\beta}}=\frac{1}{3}||E||^{2} 
\end{equation}


\end{document}